# Paying for Publication with Peer Review

## Design and Simulation of a Fee-Free, Credit-Based Open Access Journal for Information Technology

*Murat Ozer*
School of Information Technology, University of Cincinnati, Cincinnati, Ohio, United States
m.ozer@uc.edu

**Abstract**

Open access publishing has moved much of the cost of scholarly communication from readers to authors, while the labor of peer review remains unpaid and unevenly shared. This paper describes the design of Quorum: An Open Journal of Information Technology (jquorum.org), a journal that charges neither readers nor authors and uses peer review as its currency instead. Every first complete review a scholar writes on a manuscript earns one credit, and one credit pays for one submission. Each manuscript receives three double-blind reviews from reviewers matched automatically by expertise, and editorial decisions follow published rules rather than editor discretion. The paper documents the credit ledger, a bootstrapping mechanism called the genesis block, a reviewer search that widens after seven days and opens the paper to volunteers, the decision rules, and the integrity safeguards. An agent-based simulation, run for two simulated years with 40 replications per scenario, evaluates these choices. Without the genesis block, no paper was submitted in any replication. With it, a community growing from 40 to about 330 members submitted about 603 papers, and papers waited a median of 7 days for a full review panel. The day-7 escalation had a modest effect in a large community but a decisive one in a small community of about 60 members, where removing it left 48.8% of papers without a panel after 60 days, compared with 4.6% with it. Unspent credits also accumulated to about 3.7 per member after two years, which points to a policy choice the journal will face as it grows.

## 1 Introduction

Scholarly publishing depends on two contributions that are rarely paid for. Authors give away their manuscripts, and reviewers give away their time. The second contribution is large. Aczel, Szaszi and Holcombe estimated that reviewers worked over 100 million hours on journal peer review in 2020, and that the time of US-based reviewers alone was worth over 1.5 billion USD (1). The burden is also uneven. Kovanis and colleagues modeled the biomedical literature and found that 20% of researchers performed 69% to 94% of the reviews (2).

At the same time, the move to open access has shifted much of the cost of publishing from readers to authors. Many open access journals charge article processing charges (APCs) (3), which moves the access barrier from the reader side to the author side, in particular for scholars without grant funding. Journals that charge nobody, usually called diamond open access journals, exist in large numbers, but most depend on institutional subsidies and volunteer editors (4, 5). Given this context, a journal that is free for both readers and authors needs a way to secure the one resource that money usually buys indirectly: reviewer effort.

This paper describes Quorum: An Open Journal of Information Technology, which launched at jquorum.org in September 2026. Quorum treats peer review as the currency of the journal. A scholar earns one credit for each first complete review of a manuscript and spends one credit to submit a manuscript. The idea itself is not new. Fox and Petchey proposed a similar currency, which they called PubCreds, to address what they described as a tragedy of the commons in peer review (6). However, their proposal was addressed to existing journals and left the implementation open. Quorum implements the idea as a complete and automated journal, and doing so raises design questions that a proposal does not have to answer: how the economy starts when nobody has a credit, how reviewers are found when the community is small, how decisions are made without handling editors, and how the system resists gaming.

This paper makes two contributions. First, it documents the design of the journal, including the rules that govern credits, reviewer matching, decisions and integrity. Second, it evaluates the most consequential design choices with an agent-based simulation. The simulation is not a forecast of how many papers the journal will publish. It compares configurations of the same system under stated assumptions, so that the relative effect of each design choice can be seen. Please note that the author is the founding editor-in-chief of the journal, and this conflict of interest is declared at the end of the paper.

## 2 Background

### 2.1 Open Access and Its Costs

Solomon and Björk studied 1,405 journals listed in the Directory of Open Access Journals as charging APCs, which together published an estimated 103,000 articles in 2010 (3). APCs made open access financially sustainable for many publishers, but they also created a price barrier for authors. Fuchs and Sandoval introduced the term diamond open access for non-commercial journals that are free for both readers and authors (4). A study commissioned by cOAlition S surveyed 1,619 such journals and found a large but fragile sector that relies on volunteer work and small institutional budgets (5).

A fee-free journal also has to overcome suspicion. Grudniewicz and colleagues defined predatory journals as entities that prioritize self-interest at the expense of scholarship, characterized by false or misleading information, deviation from best editorial practice, a lack of transparency, and aggressive and indiscriminate solicitation (7). For this reason, a new journal cannot simply claim rigor. It has to make its rules, its reviewers' work and its funding visible.

### 2.2 Why Reviewers Decline and What Motivates Them

Tite and Schroter surveyed reviewers of five biomedical journals (551 of 890 responded) and found that conflicts with other workload, in other words a lack of time, were the most important reason for declining a review (8). Evidence on incentives is mixed. In a field experiment with referees at the Journal of Public Economics, Chetty, Saez and Sándor found that cash incentives improved review speed without crowding out intrinsic motivation, and that social incentives were especially effective among tenured professors (9). In contrast, Squazzoni, Bravo and Takács found in an experimental setting that material rewards decreased the quality and efficiency of reviewing (10). Therefore, the effect of an incentive depends on its form. A credit in Quorum is neither cash nor pure recognition. It is a right to have one's own work reviewed in return, which makes it closer to reciprocity than to payment.

### 2.3 The Reviewer Commons

Hardin described how a shared resource is depleted when each user gains from using it and bears only part of the cost (11). Peer review fits this pattern: every submission draws on the time of other scholars, while contributing reviews is voluntary. Ostrom showed that communities can govern such resources without privatization or central control when they have clear boundaries, rules that match local conditions, monitoring, graduated sanctions and accessible ways to resolve conflicts (12). Fox and Petchey took the other route and proposed to privatize the reviewer commons with PubCreds, and they discussed early-career researchers, multi-author papers, re-reviews, overdrafts and public account balances as open issues (6). As discussed in Section 3, Quorum combines both ideas: a currency, as in PubCreds, and community rules with monitoring and graduated sanctions, as in Ostrom's work.

Two further strands are relevant. Ross-Hellauer found 122 definitions of open peer review in the literature and identified publishing review reports as one of its main traits (13). Quorum publishes anonymized reports with the authors' consent. Likewise, any system that rewards a measurable action invites people to optimize the measure instead of the goal, a pattern known as Goodhart's law (14). A review credit rewards a submitted review, not a useful one, so the design has to connect the reward to quality.

## 3 Design of the Journal

Quorum is a web application with a PostgreSQL database and a scheduler that runs every five minutes. It handles submission, reviewer selection, invitations, reminders, decisions, revisions and publication without a handling editor. This section describes the rules as deployed at launch.[1]

[1] All policy values in the software are configurable. The values reported here are those deployed when the journal launched in September 2026, and the decision rules are also published on the journal's website.

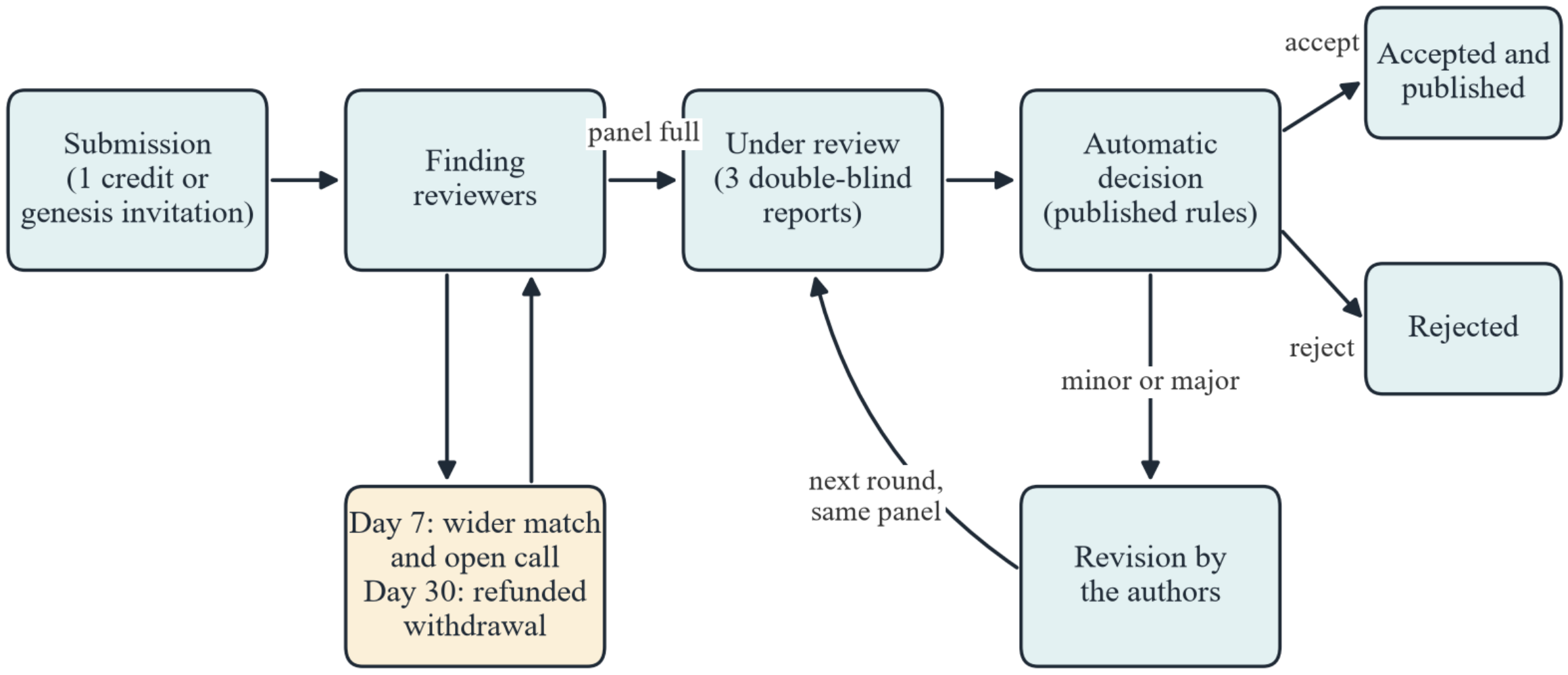

**Figure 1 Submission lifecycle in Quorum**

### 3.1 The Review Credit Ledger

Each scholar has a credit balance. Submitting a manuscript costs one credit, and the first complete review a scholar writes on a manuscript earns one credit. Re-reviews of a revised manuscript earn reputation but no additional credit. For example, a paper that goes through two review rounds creates three credits in the first round and none in the second, while its author spent one. Credits cannot be bought, sold or transferred, and donations to the journal never buy credits. An author gets the credit back when a submission is withdrawn before any reviewer has submitted a report, and also when the journal fails to complete the review panel within 30 days (Section 3.3).

The ledger is implemented as an append-only table of credit transactions. The balance stored on each account is only a cache, and it is changed only inside a database transaction together with the ledger row that explains the change. A debit is a conditional update that succeeds only when the balance covers it, so two simultaneous submissions cannot spend the same credit. A uniqueness constraint ensures that a review can earn a credit only once.

### 3.2 Bootstrapping: The Genesis Block

A credit economy that starts with no credits cannot start at all: nobody can submit, so there is nothing to review, so nobody earns a credit. Quorum solves this cold-start problem with a genesis block. The editors issue up to 20 genesis invitations to hand-picked scholars, and each invitation pays for one submission instead of a credit. Review is never waived: a genesis paper faces the same three reviewers and the same decision rules as any other paper, and a rejected genesis paper is simply rejected. The limit counts unused invitations plus genesis papers that are still alive, so a rejected or withdrawn genesis paper frees its slot. Editors cannot issue an invitation to themselves, and the founding editor's own invitation is created by the setup script. Twenty genesis papers need 60 reviews, which can mint up to 60 credits for the next wave of submissions. Section 5 shows that without this seed the simulated economy never produced a single submission.

### 3.3 Reviewer Matching and Search Escalation

Every scholar lists subject areas and expertise keywords. When a manuscript arrives, the system ranks eligible scholars by topical fit, reputation and current workload. A scholar is eligible if they accept invitations, are not paused, have fewer open assignments than their own limit (two by default), and have no conflict of interest. Conflicts are detected automatically: authors, scholars at the same institutional email domain or with the same affiliation as any author, anyone already involved with the paper, and up to three people the authors ask to exclude. Topical fit counts exact keyword matches, a match on the subject area, and the overlap of keyword tokens. A reviewer who once declined a paper as outside their area is ranked lower for that area afterwards.

On day 0 only close matches are invited, meaning scholars who share the paper's subject area or at least one exact keyword. The system invites three reviewers plus one spare. Invitations are open for four days, with a reminder 48 hours before they expire. A declined or expired invitation is replaced at once, and when the third reviewer accepts, the spare invitation is withdrawn without any penalty. If the panel is still incomplete after seven days, the search escalates. Matching widens to related areas (any shared keyword token), and these invitations are labelled as broader matches so that reviewers understand why they were chosen. At the same time, the paper appears anonymously on a page listing papers that need reviewers, where scholars in related areas can volunteer, and up to 25 of them are notified by email. On day 14 the author receives a status update, and on day 30 the author may withdraw and get the credit or genesis invitation back.

Two further rules keep the reviewer pool responsive. A scholar whose invitations expire three times in a row without any answer is paused for 30 days and can resume with one click. New members can receive invitations before they confirm their email address, but only one at a time, and accepting through the signed link in the invitation email confirms the address. This keeps the conflict checks, which rely on institutional email addresses, reliable without turning away new members.

### 3.4 Decision Rules

There is no handling editor. When the last report of a round arrives, the three recommendations are combined by the rules in Table 1, applied in order. A paper can go through at most three rounds. Reviews are due in 21 days in the first round, 14 days after a major revision and 7 days after a minor revision, with a reminder three days before the deadline. A reviewer more than five days late is removed and replaced. Authors have 60 days for a major revision and 30 days for a minor one. Accepted authors may upload a camera-ready version with names restored within seven days, after which the article is published automatically. Authors see the rationale and all reports as soon as a round is decided.

**Table 1 Decision rules, applied in order**

| Condition | Outcome |
|---|---|
| A majority of reviewers recommend rejection | Reject |
| All reviewers recommend acceptance | Accept |
| Final round (3), and nobody recommends rejection or major revision | Accept |
| Final round (3), and major concerns remain | Reject |
| Any reviewer recommends rejection or major revision | Major revision |
| Otherwise (a mix of accept and minor revision) | Minor revision |

### 3.5 Integrity Safeguards

Review is double-blind. Authors upload an anonymized PDF, and the system refuses a file whose document properties contain an author's name or whose content contains an author's email address. A step-by-step guide explains how to clear these fields in Word, LaTeX and other tools. A review counts only if it scores five criteria from 1 to 5 and contains at least 150 words of comments for the author (a third of that for re-reviews), and a review that recommends acceptance while scoring any criterion at 1 is rejected as inconsistent. After each decision, authors rate the helpfulness of every report. A rating of 5 raises the reviewer's reputation by 3 points, while a rating of 1 or 2 lowers it by 5. Editors can flag a report as low quality, which revokes its credit and costs 20 reputation points. Because reputation affects ranking, careless reviewers receive fewer invitations over time.

Editors may publish in the journal, but an editor who is an author of a submission is blocked from its editorial view, including reviewer identities and confidential comments, and from every editorial action on it. The published article carries an editorial note that says so. Editors also cannot adjust their own credit balance or issue genesis invitations to themselves. With the authors' consent, anonymized review reports are published next to the article, and reviewers may choose to sign them. In Ostrom's terms, membership with a confirmed address sets the boundary, ratings and flags provide monitoring, and reputation, pauses and revoked credits act as graduated sanctions (12). The editors themselves have a minimal role: they receive a weekly digest of stalled papers, flagged reviews and subject areas that need reviewers, and they intervene only for spam, out-of-scope work or ethics problems.

### 3.6 Publication, Indexing and Cost

Accepted articles receive a permanent identifier (for example, Quorum.2026.0001) and are published under the Creative Commons Attribution 4.0 license. Each article page carries the Highwire Press metadata that Google Scholar asks for (15), Dublin Core tags and schema.org structured data, and the PDF is served next to the abstract page. The journal also publishes a sitemap, an RSS feed and an OAI-PMH endpoint so that repositories and aggregators can harvest the archive (16). Institutions in the affiliation field are suggested from an open list of about 10,000 universities (17). The running costs, which include a server, a domain, transactional email, storage and DOI registration, are estimated at 600 to 1,200 USD per year and are covered by donations that have no effect on editorial decisions.

## 4 Simulation Study

### 4.1 Model

The simulation represents the journal day by day for two years (730 days). Scholars belong to one of six editorial sections arranged in a ring, so that each section has two neighbors. Strict matching uses the paper's own section, and broad matching adds the two neighboring sections. The community starts with 40 scholars and grows by Poisson arrivals. Each scholar has an individual probability of accepting an invitation drawn from a Beta distribution with a mean of 0.40, never answers with a probability of 0.20, and declines otherwise. A scholar with at least one credit and no paper of their own under review submits a new paper with a daily probability corresponding to three papers per year. Twenty genesis papers are submitted by founding scholars during the first 60 days.

The review process follows the deployed rules described in Section 3: three reviewers and one spare invitation, a four-day invitation window, a limit of two open assignments per reviewer, the day-7 escalation with volunteers, the day-30 withdrawal offer, and the decision rules in Table 1 with at most three rounds. Paper quality is drawn uniformly between 0.2 and 1.0, and a reviewer's recommendation is the quality plus normal noise with a standard deviation of 0.18, cut at fixed thresholds. A revision raises quality by 0.14 after a major revision and by 0.08 after a minor one. Review time follows a lognormal distribution with a median of 12 days for first reviews and 6 days for re-reviews, and 5% of reviewers miss the deadline and are replaced. After day 7, each eligible scholar volunteers for a waiting paper with a daily probability of 0.004, and half of the authors whose paper still has no panel on day 30 accept the refunded withdrawal. Table 2 lists the parameters.

**Table 2 Simulation parameters (baseline)**

| Parameter | Value |
|---|---|
| Horizon | 730 days |
| Initial scholars / arrivals | 40 / 0.40 per day |
| Editorial sections | 6, arranged in a ring |
| Genesis papers | 20 in the first 60 days |
| Invitation acceptance | Beta distribution, mean 0.40 |
| No answer | 0.20 |
| Submission rate with a credit | 3 papers per year |
| Reviewers per paper / spare invitations | 3 / 1 |
| Invitation window / reviewer load limit | 4 days / 2 open assignments |
| Review time (median) | 12 days, re-review 6 days |

| Parameter | Value |
| --- | --- |
| Missed deadlines | 5% of reviews |
| Volunteer rate after day 7 | 0.004 per eligible scholar per day |
| Withdrawal after the day-30 offer | 50% of eligible authors |
| Replications / seeds | 40 per scenario / 20260927 + r |

### 4.2 Scenarios

Seven scenarios were compared. The baseline uses the parameters in Table 2. The no-genesis scenario removes the genesis block. The no-escalation scenario removes the day-7 widening and the open call. The low-responsiveness scenario lowers the mean acceptance probability to 0.25. The slow-growth scenario lowers arrivals to 0.10 per day. Finally, a small community of 25 initial scholars with 0.05 arrivals per day was run with and without escalation, since this is closest to the conditions of a new journal. Each scenario was replicated 40 times. The simulation was written in Python with NumPy, and the code, seeds and outputs are available as described in the declarations.

## 5 Results

Table 3 summarizes the results. The first finding is categorical. Without the genesis block, no paper was submitted in any of the 40 replications, even though the community grew to about 336 members. This is the cold-start problem described in Section 3.2, and it confirms that a review-credit economy needs an initial seed of submissions to start.

**Table 3 Simulation results after two years (means of 40 replications)**

| Scenario | Members | Papers | Published | Median days to panel | 90th pct. days | No panel in 60 days | Unspent credits per member |
| --- | --- | --- | --- | --- | --- | --- | --- |
| Baseline | 330 | 603 | 318 | 7.0 | 10.2 | 0.0% | 3.7 |
| No genesis block | 336 | 0 | 0 | n/a | n/a | n/a | 0.0 |
| No escalation | 335 | 560 | 290 | 7.1 | 13.7 | 1.3% | 3.3 |
| Low responsiveness | 330 | 587 | 310 | 8.3 | 13.5 | 0.1% | 3.6 |
| Slow growth | 113 | 269 | 149 | 7.5 | 13.2 | 0.1% | 5.0 |
| Small community | 60 | 154 | 83 | 9.8 | 17.8 | 4.6% | 5.3 |
| Small community, no escalation | 61 | 81 | 25 | 7.3 | 187.7 | 48.8% | 1.7 |

In the baseline, the community grew from 40 to about 330 members, who submitted about 603 papers and published about 318 within two years (Figure 2). Submissions started slowly with the genesis papers and accelerated as reviews minted credits. The median time from submission to a full review panel was 7 days, the 90th percentile was 10.2 days, and the median time to the first decision was 25.1 days. Please note that 57.2% of papers were still short of a full panel on their seventh day. Because invitations stay open for four days and an unanswered invitation is replaced only when it expires, a typical panel formed around the seventh day, which means the escalation stage was reached by most papers.

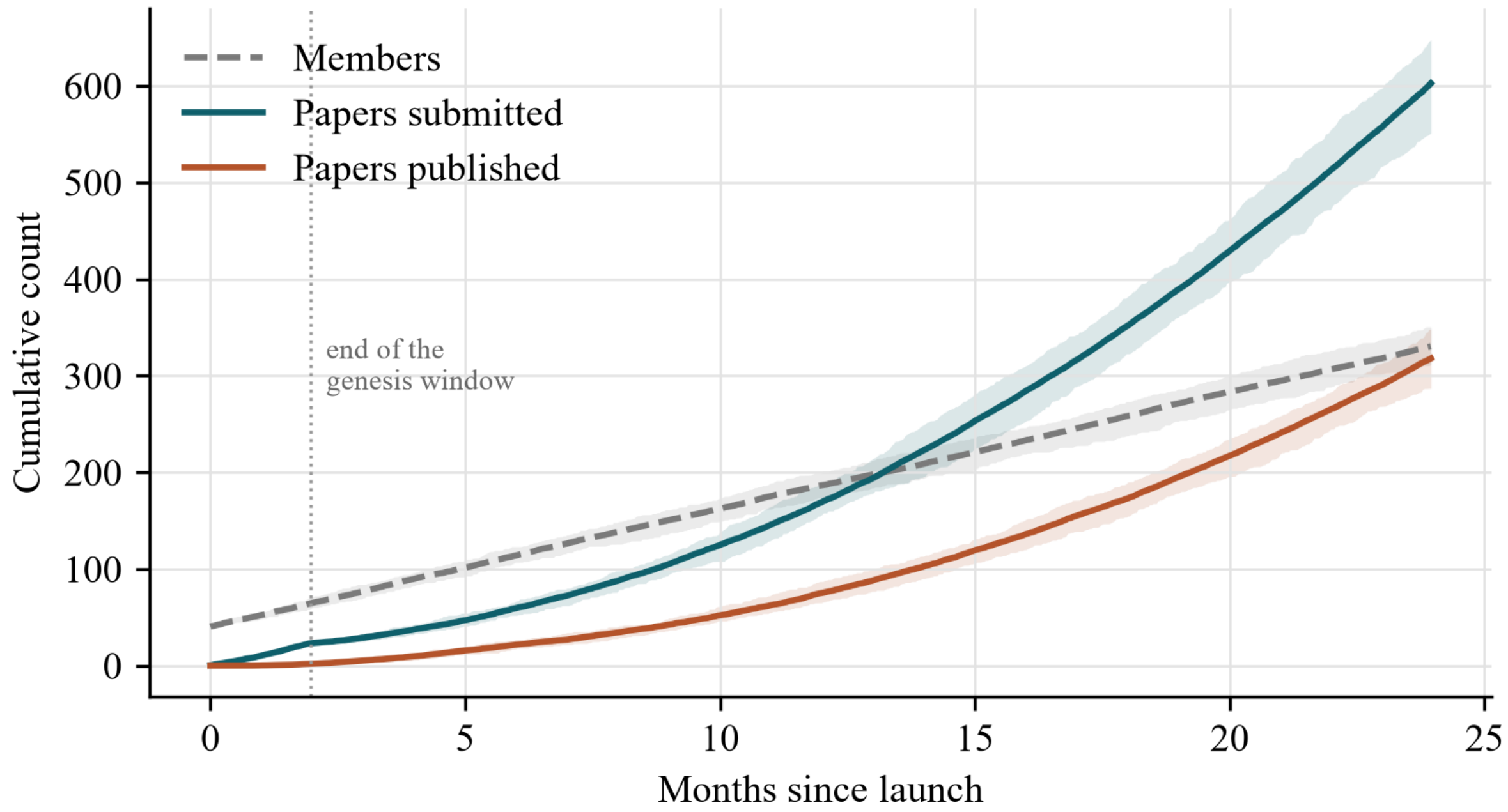

**Figure 2 Baseline growth of members, submissions and publications**

The day-7 escalation had a modest effect in the baseline community. Without it, the 90th percentile of the time to a full panel rose from 10.2 to 13.7 days, 1.3% of papers had no panel after 60 days, and about 4.5 papers per run were withdrawn after the day-30 offer. Lower responsiveness had a similar, modest effect: with a mean acceptance probability of 0.25, the median wait rose to 8.3 days and 82.1% of papers reached the escalation stage, but almost all papers still found a panel (Figure 4).

The picture changes in a small community (Figure 3). With about 60 members after two years, the journal published about 82.5 papers with escalation and only 25.4 without it, a ratio of 3.2 to 1. Without escalation, 48.8% of papers had no panel after 60 days, compared with 4.6% with it, and the number of papers waiting for reviewers kept rising throughout the two years instead of settling. Among papers that did find a panel, the 90th percentile wait was 188 days without escalation and 17.8 days with it. Likewise, the share of members who wrote at least one review fell from 95% to 66% without escalation, because strict matching kept most of the community out of reach of most papers.

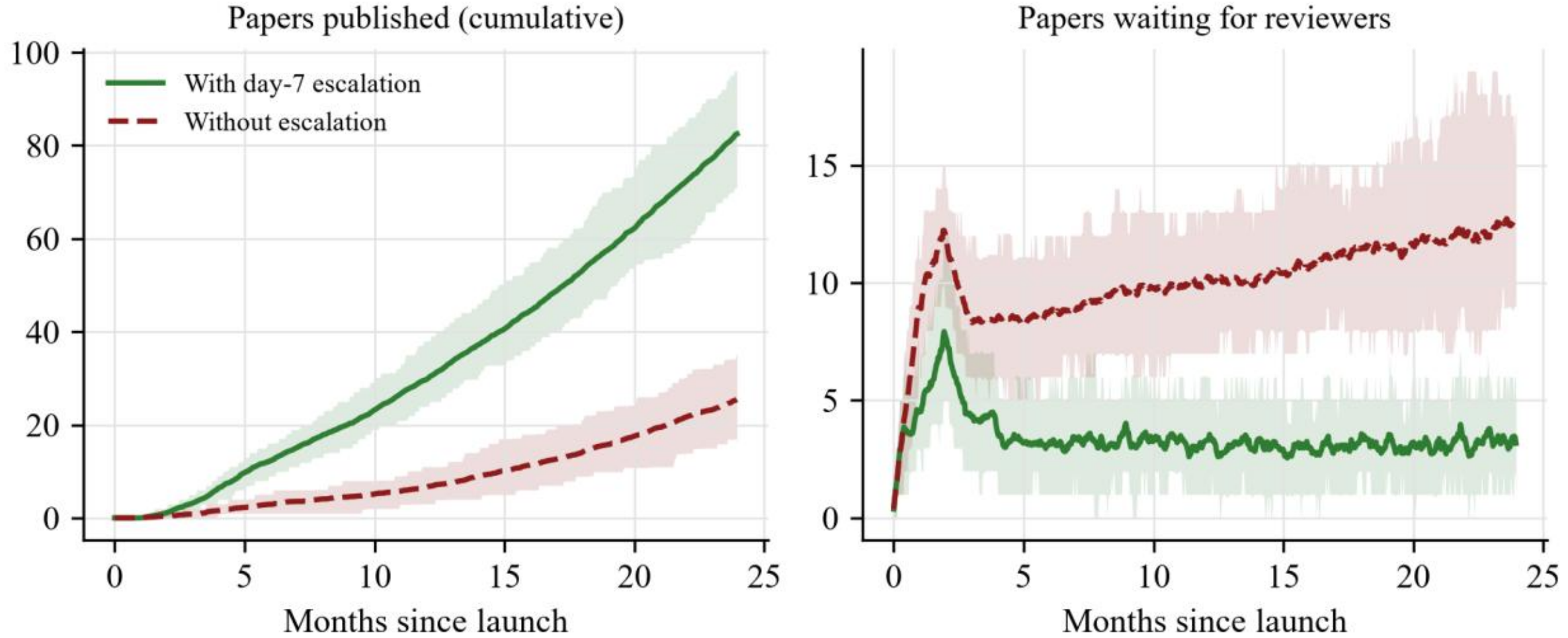

**Figure 3 Small community with and without the day-7 escalation**

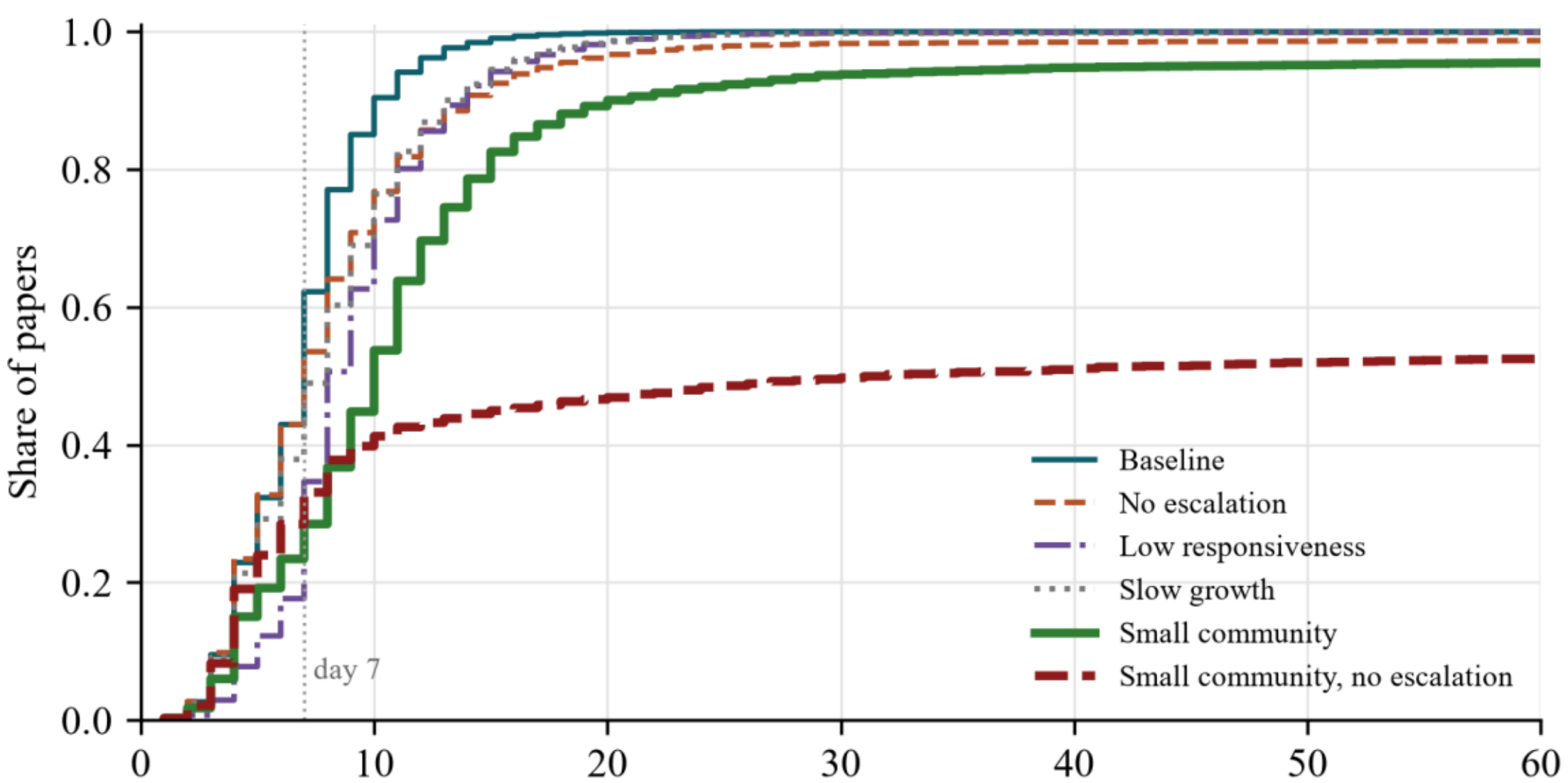

**Figure 4 Time from submission to a full review panel**

Credits accumulated in every scenario that produced papers (Figure 5). In the baseline, reviews minted about 1800 credits and submissions spent about 583, leaving about 3.7 unspent credits per member after two years. Accumulation was faster in the slow-growth and small-community scenarios (5.0 and 5.3 per member), where fewer new members arrived with empty balances. The work of reviewing was spread widely. In the baseline, 93% of members wrote at least one review, and the most active 20% of reviewers wrote 38% of all reviews.

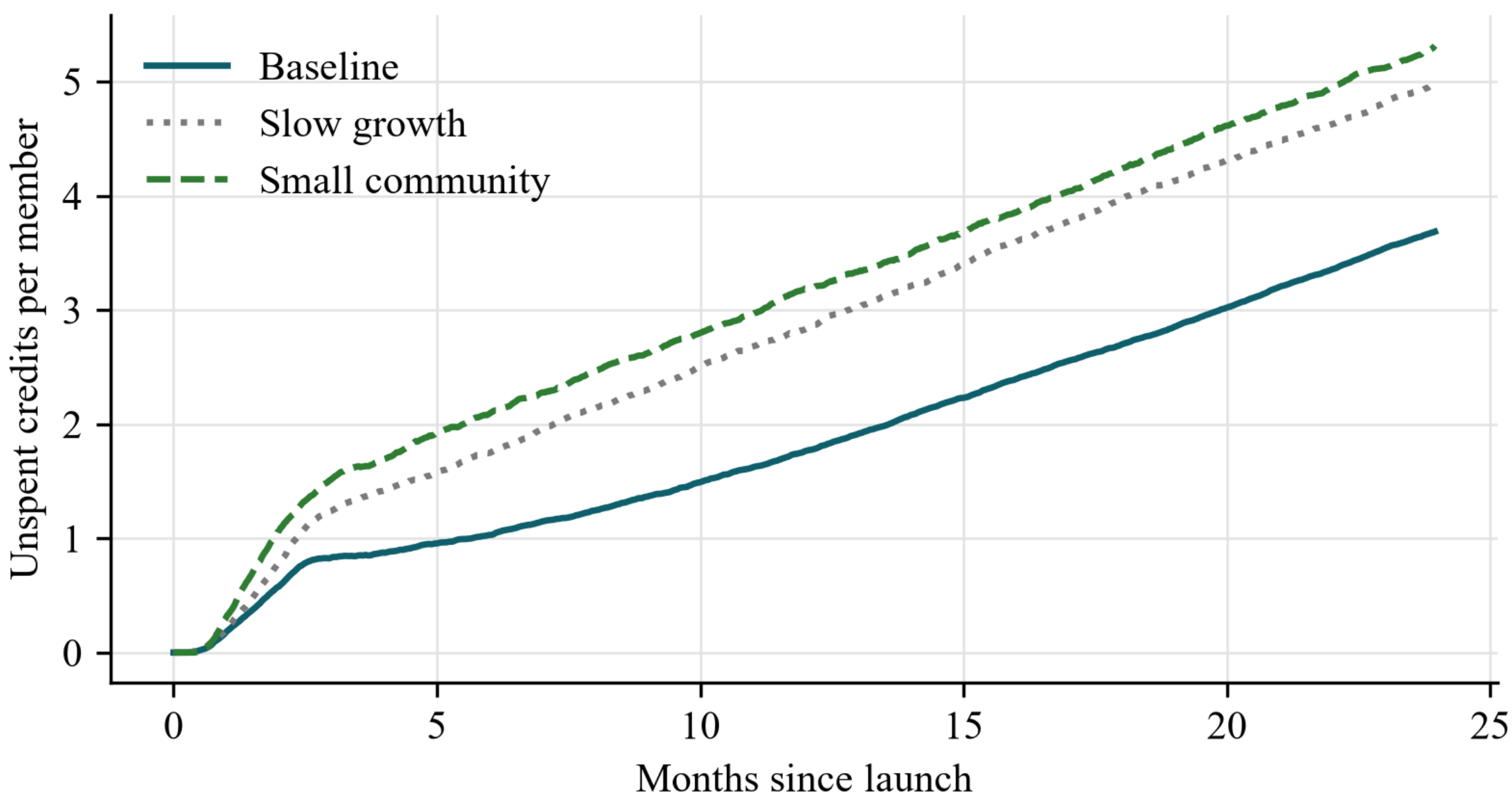

**Figure 5 Unspent credits per member over time**

## 6 Discussion

### 6.1 Main Findings

The simulation supports three design conclusions. First, the genesis block is not optional. A review-credit economy with no initial credits has no activity at all, and the size of the seed matters less than its existence, because the first reviewed papers mint the credits for the next ones. Second, escalation is what keeps a small journal alive. In a large community, strict matching finds reviewers for almost every paper and escalation mostly shortens the tail of the waiting time. In a community of about 60 members, strict matching alone leaves half of the papers without reviewers, and widening the match together with an open call for volunteers roughly triples the number of published papers. Given this context, the journal should keep escalation on during its first years, when its community is closest to the small-community scenario. Third, the timing of escalation interacts with the invitation window. Because a panel typically formed around the seventh day, a shorter invitation window or a larger spare buffer might shorten waiting times further, and this is a parameter the journal can tune with real data.

### 6.2 Credit Accumulation

Each paper can mint up to three credits in its first round and costs its author one, so credits accumulate whenever the review process works. This is intended in the early phase, because it lets new reviewers become authors quickly. However, a large surplus weakens the incentive the currency is meant to create. A scholar with five unspent credits has little reason to accept the next invitation. Readers might ask why the journal does not simply mint one credit per paper instead of three. The answer is that every paper needs three reviewers, and each of them should be paid for their work. The more promising options act on balances rather than on the price of a review: a cap on unspent credits (for example, three), credits that expire after two years, or a rule that keeps invitations flowing to scholars with large balances first. None of these is implemented at launch, and the journal will choose among them once real balances are observed.

### 6.3 Gaming and Review Quality

A credit rewards a submitted review, so the design has to prevent low-effort reviews from being profitable, which is the risk that Goodhart's law describes (14). Quorum addresses it at several points: minimum length and consistency checks at submission, author ratings after the decision, editor flags that revoke credit, reputation that decides who is invited first, and published reports that expose superficial reviews to public view. The field evidence that material rewards can reduce review quality (10) is a reason to keep the reward reciprocal and non-monetary, which Quorum does. Collusion, such as two scholars rating each other's reviews highly, is not modeled in the simulation. Even though this is a limitation, the double-blind design makes targeted collusion hard, because authors do not know who reviewed their paper when they rate the report.

### 6.4 Limitations

The simulation parameters are assumptions, not estimates. The journal launched shortly before this paper was written, so no empirical data on acceptance rates, review times or submission rates were available. For this reason, the absolute numbers in Table 3 should be read only within the model. However, the comparisons between scenarios, which are the purpose of the simulation, are less sensitive to the exact values because every scenario shares the same assumptions. The model also simplifies scholarly behavior. Each scholar belongs to one section, submission depends only on holding a credit, paper quality is a single number, and conflicts of interest, which shrink the reviewer pool, are not modeled. The last simplification makes the results optimistic for reviewer availability, which strengthens rather than weakens the conclusion that escalation matters in small communities. Finally, the comparison of review concentration with the biomedical literature (2) illustrates how a load limit spreads work, but the model's reviewers differ only in their willingness to accept, so it should not be read as a prediction of real concentration.

## 7 Conclusion

Quorum tests whether a journal can be free for readers and authors by paying for publication with peer review. The design turns the reviewer commons into a small economy with a ledger, a seed, automated matching and community rules for quality. The simulation shows that the seed is necessary for the economy to start, that widening the reviewer search after seven days is essential while the community is small, and that credit accumulation will need a policy response as the journal grows. The journal records every decision, credit and review in an auditable form, so these findings can be tested against real data in follow-up work.

## Declarations

Conflict of interest: The author is the founding editor-in-chief of Quorum: An Open Journal of Information Technology, the journal described in this paper.

Funding: This research did not receive any specific grant from funding agencies in the public, commercial, or not-for-profit sectors.

Data and code availability: The simulation code, the scenario definitions with their random seeds, and the complete outputs are available from the author and will be deposited in a public repository (https://github.com/ozermm/quorum-credit-economy-sim).

Ethics: No human participants or personal data were used. The simulation uses generated data only.

Use of AI tools: Generative AI (Claude, Anthropic) assisted in writing the simulation code. The author reviewed and edited the analysis and remains responsible for the manuscript.